\PassOptionsToClass{nojss}{jss}
\documentclass[article]{jss}

\usepackage{orcidlink,thumbpdf,lmodern}

\usepackage{amsmath,amssymb,booktabs,array}

\newcommand{\fct}[1]{\code{#1()}}
\DeclareMathOperator{\logit}{logit}

\author{Tom M. Palmer~\orcidlink{0000-0003-4655-4511}\\MRC Integrative\\Epidemiology Unit\\University of Bristol
   \And Nicolai Vitt~\orcidlink{0000-0002-5988-9372}\\MRC Integrative\\Epidemiology Unit\\University of Bristol
   \And Eleanor Sanderson~\orcidlink{0000-0001-5188-5775}\\MRC Integrative\\Epidemiology Unit\\University of Bristol}
\Plainauthor{Tom M. Palmer, Nicolai Vitt, Eleanor Sanderson}

\title{One-Sample {M}endelian Randomization and Instrumental Variable Analyses in \proglang{R} and \proglang{Stata}: The \pkg{OneSampleMR} and \pkg{ivonesamplemr} Packages}
\Plaintitle{One-Sample Mendelian Randomization and Instrumental Variable Analyses in R and Stata: The OneSampleMR and ivonesamplemr Packages}
\Shorttitle{One-Sample {M}endelian Randomization in \proglang{R} and \proglang{Stata}}

\Abstract{
  Mendelian randomization (MR) uses genetic variants as instrumental variables
  (IVs) to estimate the causal effect of an exposure on an outcome. When
  individual-level data on the instruments, exposure, and outcome are available
  in a single sample (one-sample MR), a range of IV estimators can be used
  beyond the standard two-stage least squares estimator, including structural
  mean models (SMMs) and two-stage predictor substitution (TSPS) and two-stage
  residual inclusion (TSRI) estimators with nonlinear second stage models.
  These estimators are important when the outcome is binary, where
  investigators typically wish to report causal risk ratios or causal odds
  ratios. However, implementations of these estimators are scattered across
  software packages and are often unavailable with standard errors that
  account for the estimation of the first stage. We describe two packages
  which implement these estimators: \pkg{OneSampleMR} for \proglang{R} and
  \pkg{ivonesamplemr} for \proglang{Stata}. Both packages fit the
  multiplicative and logistic SMMs and TSPS and TSRI estimators with
  identity, log-additive, log-multiplicative, and logit links using the
  generalized method of moments (GMM), so that the standard errors account
  for the uncertainty in both stages of estimation. Both packages also provide
  the Sanderson-Windmeijer conditional $F$-statistic for assessing instrument
  strength in models with multiple exposures, and moving window and
  within-quantile estimation for exploring nonlinear exposure-outcome
  relationships. We illustrate both packages using
  the same simulated dataset and show that they give the same estimates as
  each other and as existing implementations.
}

\Plainkeywords{Mendelian randomization, instrumental variables, structural mean models, generalized method of moments, R, Stata}

\Address{
  Tom M. Palmer\\
  MRC Integrative Epidemiology Unit\\
  Population Health Sciences\\
  Bristol Medical School\\
  University of Bristol\\
  Augustine's Courtyard, Orchard Lane\\
  Bristol, BS1 5DS, United Kingdom\\
  E-mail: \email{tom.palmer@bristol.ac.uk}\\
  URL: \url{https://remlapmot.github.io/}
}

\begin{document}

%% -- Introduction -------------------------------------------------------------

\section{Introduction} \label{sec:intro}

Mendelian randomization (MR) is an approach to estimating the causal effect of
a modifiable exposure on an outcome using genetic variants as instrumental
variables (IVs) \citep{DaveySmith+Ebrahim:2003, Lawlor+Harbord+Sterne:2008,
Didelez+Sheehan:2007, Sanderson+Glymour+Holmes:2022}. Because genotypes are
allocated at gamete cell formation and then conception, they are expected to be largely independent of the
environmental and behavioral factors that commonly confound observational
associations. Under the IV assumptions, described in Section~\ref{sec:models},
the association of the genetic variants with the outcome can be used to make
inferences about the causal effect of the exposure on the outcome.

MR analyses are conducted in one of two designs. In \emph{two-sample} MR, the
gene-exposure and gene-outcome associations are estimated in different samples
and only summary statistics are required. Software for two-sample MR is
plentiful, for example the \proglang{R} packages \pkg{MendelianRandomization}
\citep{Yavorska+Burgess:2017} and \pkg{TwoSampleMR}
\citep{Hemani+Zheng+Elsworth:2018}, and the \proglang{Stata} package
\pkg{mrrobust} \citep{Spiller+Davies+Palmer:2019}. In \emph{one-sample} MR,
individual-level data on the genetic variants, exposure, and outcome are
available for the same participants, for example in a cohort study or
biobank. This design permits the use of the full range of IV estimators
developed in biostatistics and econometrics.

For a continuous outcome, the standard one-sample analysis is linear IV
estimation, such as two-stage least squares (2SLS), which is widely available.
However, many outcomes in epidemiology are binary, such as disease status, and
investigators usually wish to report a causal risk ratio or a causal odds ratio
rather than a causal risk difference. Several estimators have been proposed for
this setting \citep{Palmer+Sterne+Harbord:2011, Didelez+Meng+Sheehan:2010}.
These include the multiplicative structural mean model
\citep[MSMM;][]{Robins:1989, Robins:1994, Hernan+Robins:2006}, which estimates a
causal risk ratio, the logistic structural mean model
\citep[LSMM;][]{Vansteelandt+Goetghebeur:2003, Vansteelandt+Bowden+Babanezhad:2011},
which estimates a causal odds ratio, and two-stage predictor substitution (TSPS)
and two-stage residual inclusion (TSRI) estimators
\citep{Terza+Basu+Rathouz:2008} with nonlinear second stage models.
\citet{Clarke+Palmer+Windmeijer:2015} showed how SMMs can be estimated with
multiple instruments using the generalized method of moments
\citep[GMM;][]{Hansen:1982}. When there are more instruments than exposures,
GMM also provides a test of the overidentifying restrictions, which can detect
some violations of the independence and exclusion restriction assumptions
(Section~\ref{sec:models}) only insofar as they cause the instruments to disagree about the causal effect. GMM estimation of TSPS and TSRI estimators, in which the
first and second stage estimating equations are stacked, gives standard errors
which correctly account for the estimation of the first stage, which naive
two-step estimation does not.

In addition, assessing the strength of the instruments is an important part of
any IV analysis because weak instruments bias IV estimators towards the
confounded observational association. With a single exposure, the first stage
$F$-statistic is typically reported \citep{Stock+Yogo:2005}. With multiple
endogenous exposures, as in multivariable MR, the individual first stage $F$-statistics
are not informative about whether each exposure is separately identified, and
the conditional $F$-statistic of \citet{Sanderson+Windmeijer:2016} should be
used instead.

Implementations of these methods exist but are scattered. In \proglang{R}
\citep{R}, linear IV estimation is available in \pkg{ivreg} \citep{ivreg},
\pkg{AER} \citep{Kleiber+Zeileis:2008}, \pkg{estimatr} \citep{estimatr}, and
\pkg{fixest} \citep{fixest}. The \pkg{ivtools} package
\citep{Sjolander+Martinussen:2019} provides G-estimation of SMMs and two-stage
estimators, along with methods for survival outcomes and nonparametric bounds.
A user can also code the relevant moment conditions themselves using the
\pkg{gmm} package \citep{Chausse:2010}. In \proglang{Stata} \citep{Stata}, the
official \code{ivpoisson} command \citep{Stata:ivpoisson} fits an estimator
that is equivalent to the MSMM \citep{Mullahy:1997, Windmeijer+SantosSilva:1997,
Palmer+Sterne+Harbord:2011}, as does the earlier community-contributed
\code{ivpois} \citep{Nichols:2007}, and \code{ivreg2}
\citep{Baum+Schaffer+Stillman:2007} reports Sanderson-Windmeijer $F$-statistics.
However, as far as we are aware, there is no other software that fits TSPS and
TSRI estimators with log-multiplicative second stage models, nor any package
that provides the same set of one-sample MR estimators with the same
parameterizations in both \proglang{R} and \proglang{Stata}. Moreover, the
existing implementations are spread across packages with different interfaces
and terminology, several of which require the user to fit and combine component
models. Having the main one-sample estimators in one place, with a common
interface and documentation aimed at MR analysts, in the way that
\pkg{TwoSampleMR} and \pkg{MendelianRandomization} do for two-sample MR, makes
them easier to apply and to compare.

In this article we describe two packages that fill this gap.
\pkg{OneSampleMR} \citep{OneSampleMR} is an \proglang{R} package available from
the Comprehensive \proglang{R} Archive Network (CRAN) at
\url{https://CRAN.R-project.org/package=OneSampleMR}. \pkg{ivonesamplemr}
\citep{ivonesamplemr} is a \proglang{Stata} package available from the
Statistical Software Components (SSC) archive, with its development version
available from its GitHub repository,
\url{https://github.com/remlapmot/ivonesamplemr}. Both are released under the
GNU General Public License version 3. The packages were developed in parallel
and provide the same functionality, so that analysts who work in either
language, or collaborators who work in different languages, can fit the same
models and obtain the same results.
The \proglang{Stata} package has been used in applied MR analyses, for example
\citet{MadleyDowd+Dardani+Wootton:2022}.

The rest of the article is organized as follows. Section~\ref{sec:models}
describes the models and estimators. Section~\ref{sec:software} compares the
packages with existing software. Section~\ref{sec:data} describes the
simulated dataset used in the illustrations. Sections~\ref{sec:r} and
\ref{sec:stata} illustrate \pkg{OneSampleMR} and \pkg{ivonesamplemr},
respectively. Section~\ref{sec:agreement} compares the estimates across
implementations, and Section~\ref{sec:summary} concludes with a discussion.

%% -- Models and estimators ----------------------------------------------------

\section{Models and estimators} \label{sec:models}

\subsection{Notation and assumptions}

Let $Y$ denote the outcome, $X$ the exposure (or a vector of $K$ exposures),
which is endogenous, $Z$ a vector of $L$ instruments (typically genotypes coded as the number of effect
alleles, or genetic risk scores combining several variants), $W$ a vector of measured covariates, and $U$ the unmeasured
confounders of the exposure-outcome relationship. Let $Y(x)$ denote the
potential outcome that would be observed if the exposure were set to $x$, so
that $Y(0)$ is the exposure-free potential outcome. The instruments must
satisfy three core IV assumptions \citep{Didelez+Sheehan:2007,
Hernan+Robins:2006}:
\begin{enumerate}
  \item[IV1.] Relevance: $Z$ is associated with $X$.
  \item[IV2.] Independence: $Z$ is independent of $U$ (conditional on $W$).
  \item[IV3.] Exclusion restriction: $Z$ is independent of $Y$ given $X$, $U$
    (and $W$), that is, $Z$ affects $Y$ only through $X$.
\end{enumerate}
These assumptions alone do not point identify the causal effect, and each
estimator below makes an additional structural or parametric assumption, such
as no effect modification of the causal effect by the instruments.

\subsection{Structural mean models}

SMMs \citep{Robins:1989, Robins:1994} specify a model for the contrast between
the mean of the observed outcome and the mean of the exposure-free potential
outcome among those with the same exposure and instrument values.

The \emph{additive} SMM (ASMM) is
\begin{equation*}
  \E(Y \mid X, Z) - \E\{Y(0) \mid X, Z\} = \psi X.
\end{equation*}
Under the IV assumptions, $Y - \psi X$ is mean independent of $Z$, which gives
the estimating equations $\E\{(1, Z^\top)^\top (Y - \beta_0 - \psi X)\} = 0$,
where $\beta_0 = \E\{Y(0)\}$. These are the estimating equations of linear IV estimators, and so the ASMM is
fit by 2SLS or linear GMM. For a binary outcome, $\psi$ is a causal risk
difference.

The \emph{multiplicative} SMM (MSMM) is
\begin{equation*}
  \log \E(Y \mid X, Z) - \log \E\{Y(0) \mid X, Z\} = \psi X,
\end{equation*}
so that $\exp(\psi)$ is a causal risk ratio for a binary outcome. Under the IV
assumptions $Y\exp(-\psi X)$ is mean independent of $Z$, giving the moment
conditions
\begin{equation}
  \E\left[ (1, Z^\top)^\top \{ Y \exp(-\psi X) - \beta_0 \} \right] = 0,
  \label{eq:msmm}
\end{equation}
where again $\beta_0 = \E\{Y(0)\}$. \citet{Clarke+Palmer+Windmeijer:2015} described
GMM estimation of \eqref{eq:msmm} with multiple instruments, and an alternative
parameterization,
\begin{equation}
  \E\left[ (1, Z^\top)^\top \{ Y \exp(-\beta_0^* - \psi X) - 1 \} \right] = 0,
  \label{eq:msmmalt}
\end{equation}
which gives the same moment conditions as the estimator of
\citet{Mullahy:1997} implemented in \proglang{Stata}'s \code{ivpoisson}
command with the \code{multiplicative} option. Here $\exp(\beta_0^*) =
\E\{Y(0)\}$. \citet{Dukes+Vansteelandt:2018} noted that these moment conditions
are equivalent to those of a gamma regression. When $X$ and $Y$ are binary,
\citet{Clarke+Palmer+Windmeijer:2015} also showed that the MSMM can be fitted
with 2SLS using transformed variables.

The \emph{logistic} SMM (LSMM) is
\begin{equation*}
  \logit \E(Y \mid X, Z) - \logit \E\{Y(0) \mid X, Z\} = \psi X,
\end{equation*}
so that $\exp(\psi)$ is a causal odds ratio among the exposed. Unlike the ASMM
and MSMM, the LSMM is not identified from the IV assumptions alone; it also
requires a model for the observed outcome mean, the so-called association
model, for example $\logit \E(Y \mid X, Z) = \alpha^\top (1, X, Z^\top)^\top$.
The double logistic SMM of \citet{Vansteelandt+Goetghebeur:2003} jointly fits
the association model and the causal model. With multiple instruments the
estimating equations can again be solved by GMM
\citep{Clarke+Palmer+Windmeijer:2015}.

When there are more instruments than exposures, $L > K$, the SMMs are
overidentified, and Hansen's $J$-statistic tests the overidentifying
restrictions, that is, whether the moment conditions implied by the different
instruments are compatible with each other \citep{Hansen:1982}. Under the null
hypothesis that all the instruments are valid, $J$ has an asymptotic $\chi^2$
distribution with $L - K$ degrees of freedom. Rejection indicates that at least
one instrument violates the independence or exclusion restriction assumptions,
or that the model is misspecified, for example because the causal effect
differs between subgroups defined by the instruments. The test has low power
when all the instruments are invalid in a similar way, so a nonsignificant
$J$-statistic does not establish that the instruments are valid.

\subsection{Two-stage predictor substitution and residual inclusion}

TSPS and TSRI estimators \citep{Terza+Basu+Rathouz:2008} proceed by fitting a
first stage linear regression of the exposure on the instruments and
covariates,
\begin{equation*}
  X = \alpha_0 + \alpha_Z^\top Z + \alpha_W^\top W + e.
\end{equation*}
TSPS then fits a second stage model, with link function $g$, of the outcome on
the fitted values from the first stage, $\hat{X}$,
\begin{equation*}
  g\{\E(Y \mid \hat{X}, W)\} = \beta_0 + \beta_X \hat{X} + \beta_W^\top W,
\end{equation*}
while TSRI fits the second stage model of the outcome on the exposure and the
first stage residuals, $\hat{e} = X - \hat{X}$,
\begin{equation*}
  g\{\E(Y \mid X, \hat{e}, W)\} = \beta_0 + \beta_X X + \beta_e \hat{e} + \beta_W^\top W.
\end{equation*}
TSRI is a special case of a control function estimator. Both \pkg{OneSampleMR}
and \pkg{ivonesamplemr} offer four link functions for the second stage model:
\begin{itemize}
  \item \code{identity}: linear regression, with second stage residual
    $Y - \eta$, where $\eta$ is the second stage linear predictor; $\beta_X$
    is a causal risk difference for a binary outcome.
  \item \code{logadd}: log-additive, corresponding to Poisson or log-binomial
    regression, with residual $Y - \exp(\eta)$; $\exp(\beta_X)$ is a causal
    risk ratio for a binary outcome.
  \item \code{logmult}: log-multiplicative, corresponding to gamma
    regression, with residual $Y\exp(-\eta) - 1$; $\exp(\beta_X)$ is a causal
    risk ratio for a binary outcome \citep{Dukes+Vansteelandt:2018}.
  \item \code{logit}: logistic regression, with residual
    $Y - \logit^{-1}(\eta)$; $\exp(\beta_X)$ is a causal odds ratio for a
    binary outcome.
\end{itemize}

In both \pkg{OneSampleMR} and \pkg{ivonesamplemr} the first stage estimating equations, $(1, Z^\top,
W^\top)^\top e$, and the second stage estimating equations, the second stage
residual multiplied by the second stage regressors, are stacked and solved
jointly by GMM. Any number of instruments can be used, because the instruments
enter only the first stage regression. The stacked system therefore always has
as many estimating equations as parameters, that is, it is exactly identified,
whatever the number of instruments. As a result the point estimates equal
those of the sequential two-step procedure, the GMM standard errors account for
the uncertainty in the first stage estimates, and, unlike for the SMMs, no
overidentification test is available. Both \pkg{OneSampleMR} and
\pkg{ivonesamplemr} currently allow only a single exposure in the TSPS and TSRI
estimators.

With the identity link, TSPS is numerically identical to 2SLS. With nonlinear
links the TSPS and TSRI estimators are consistent for the causal effect only
under additional assumptions on the first stage and on the distribution of the
unmeasured confounders. In particular, with the logit link, the TSPS and TSRI
estimands are generally not the population or exposed causal odds ratio,
because of the noncollapsibility of the odds ratio
\citep{Palmer+Sterne+Harbord:2011, Burgess:2013, Didelez+Meng+Sheehan:2010}.
The log-additive and log-multiplicative links avoid this problem for risk
ratios. These estimators therefore complement, rather than replace, the SMMs.

\subsection{Instrument strength with multiple exposures}

In a model with $K$ exposures and $L \geq K$ instruments, the
\citet{Sanderson+Windmeijer:2016} conditional $F$-statistic for exposure $j$
tests whether exposure $j$ is weakly identified given the other exposures. It
is computed from a regression of exposure $j$ on the instruments after
partialling out the linear projection of the other exposures on the
instruments, and has $L - K + 1$ numerator degrees of freedom. When $K = 1$ it
reduces to the usual first stage $F$-statistic for the excluded instruments.
The conditional $F$-statistics can be compared with the
\citet{Stock+Yogo:2005} critical values for a single exposure with $L - K + 1$
instruments ($k_z - 1$ instruments in the two-exposure case considered by
\citealp{Sanderson+Windmeijer:2016}, where $k_z$ is the number of
instruments).

\subsection{Nonlinear exposure-outcome relationships}

The estimators above assume a linear (on the scale of the link function)
exposure-outcome relationship. \citet{Burgess+Davies+Thompson:2014} described
two approaches to exploring departures from linearity: estimating the causal
effect within strata defined by quantiles of the first stage residuals
(equivalently, of the IV-free exposure), and a moving (sliding) window approach
in which the estimator is applied to successive overlapping windows of
participants ordered by the IV-free exposure. Comparing the estimates across
strata or windows indicates whether the causal effect varies with the level of
the exposure. These approaches are implemented in both packages, by the
\fct{ivxtile} and \fct{ivmw} functions in \pkg{OneSampleMR} and the
\code{ivxtile} and \code{ivmw} prefix commands in \pkg{ivonesamplemr}
(Sections~\ref{sec:r} and~\ref{sec:stata}).
They rely on an additional assumption, that the effect of the instruments on the
exposure is the same in every stratum or window (the constant genetic effect
assumption). When this assumption is violated, for example when the genetic
effect on the exposure varies with the level of the exposure, the
stratum-specific estimates can be biased and can suggest a nonlinear
relationship where none exists \citep{Burgess:2023, Hamilton+Hughes+Spiller:2024}.

%% -- Existing software --------------------------------------------------------

\section{Comparison with existing software} \label{sec:software}

Table~\ref{tab:software} summarizes the estimators available in
\pkg{OneSampleMR}, \pkg{ivonesamplemr}, and related software.

\begin{table}[t!]
\centering
\small
\setlength{\tabcolsep}{4pt}
\begin{tabular}{lcccccc}
\toprule
 & \multicolumn{3}{c}{\proglang{R}} & \multicolumn{3}{c}{\proglang{Stata}} \\
\cmidrule(lr){2-4} \cmidrule(lr){5-7}
Estimator & \pkg{OneSampleMR} & \pkg{ivtools} & \pkg{ivreg} & \pkg{ivonesamplemr} & Official & \code{ivreg2} \\
\midrule
ASMM (linear IV)              & via \pkg{ivreg} & \checkmark  & \checkmark & via \code{ivregress} & \checkmark & \checkmark \\
MSMM, moment conditions \eqref{eq:msmm}    & \checkmark   & --$^a$      & --         & \checkmark & --         & -- \\
MSMM, moment conditions \eqref{eq:msmmalt} & \checkmark   & --          & --         & \checkmark & \checkmark$^b$ & -- \\
MSMM, 2SLS methods            & \checkmark   & --          & --         & \checkmark & --         & -- \\
LSMM                          & \checkmark   & \checkmark  & --         & \checkmark & --         & -- \\
TSPS, identity/log/logit link & \checkmark   & \checkmark  & --         & \checkmark & --         & -- \\
TSPS, log-multiplicative link & \checkmark   & --          & --         & \checkmark & --         & -- \\
TSRI, identity/log/logit link & \checkmark   & \checkmark  & --         & \checkmark & --         & -- \\
TSRI, log-multiplicative link & \checkmark   & --          & --         & \checkmark & --         & -- \\
Sanderson-Windmeijer $F$      & \checkmark   & --          & --         & \checkmark & --         & \checkmark \\
Overidentification test       & \checkmark   & --          & \checkmark & \checkmark & \checkmark & \checkmark \\
Moving window / quantiles     & \checkmark   & --          & --         & \checkmark & --         & -- \\
\bottomrule
\end{tabular}
\caption{\label{tab:software} Estimators available in \pkg{OneSampleMR},
\pkg{ivonesamplemr}, and related software, for multiple instruments.
$^a$\pkg{ivtools} G-estimates the MSMM with a single instrument. $^b$\code{ivpoisson gmm} with the
\code{multiplicative} option, which uses the moment conditions in
\eqref{eq:msmmalt}.}
\end{table}

The package with the most overlap with \pkg{OneSampleMR} is \pkg{ivtools}
\citep{Sjolander+Martinussen:2019}. Its \fct{ivglm} function fits two-stage
estimators (with the \code{ctrl} argument selecting between predictor
substitution and residual inclusion) and G-estimators of SMMs with identity,
log, and logit links, and \pkg{ivtools} additionally provides IV estimators
for survival outcomes and nonparametric bounds. Its interface requires the
user to fit and supply the component models, for example a model for the
instrument for G-estimation. In contrast, \pkg{OneSampleMR} uses the
three-part formula interface, \code{y ~ x + w | z + w}, of \fct{ivreg} from
\pkg{ivreg} and \pkg{AER}, fits the MSMM and LSMM with multiple instruments by
GMM with an overidentification test, offers the log-multiplicative link for TSPS and
TSRI, computes the Sanderson-Windmeijer conditional $F$-statistic for IV
models fitted by \pkg{ivreg}, \pkg{AER}, \pkg{estimatr}, and \pkg{fixest},
and provides within-quantile and moving window estimation for any of its
estimators.
This makes \pkg{OneSampleMR} simple to use, because each model is specified in
a single function call in the same way as a linear IV model, rather than by
fitting and combining several component models.

In \proglang{Stata}, \code{ivmsmm} fits the MSMM by default as a wrapper for
\code{ivpoisson gmm} with the \code{multiplicative} option, which uses the
moment conditions in \eqref{eq:msmmalt}, and with its \code{estmethod()}
option also fits the moment conditions in \eqref{eq:msmm} and the two-stage
least squares methods of \citet{Clarke+Palmer+Windmeijer:2015}. The
\code{ivfsw} command computes the Sanderson-Windmeijer conditional
$F$-statistics after \code{ivregress} or \code{ivreg2}, giving the same values
as \code{ivreg2} with its \code{first} option. The remaining
\pkg{ivonesamplemr} commands, \code{ivlsmm}, \code{ivtsps}, \code{ivtsri},
\code{ivmw}, and \code{ivxtile}, provide estimators that are not otherwise
available in official \proglang{Stata} or, to our knowledge, in other
community-contributed packages, other than via user-written moment conditions
for \proglang{Stata}'s \code{gmm} command.

The most distinctive feature of the two packages is that they provide the same
estimators, with the same parameterizations and link function names, so that
the same analysis can be reproduced in either language. We demonstrate this in
Section~\ref{sec:agreement}. In addition, both packages build on the syntax of
the most widely used linear IV commands, \fct{ivreg} in \proglang{R} and
\code{ivregress} in \proglang{Stata}, so users who are familiar with linear IV
estimation can fit the other estimators by changing only the function or
command name and, for TSPS and TSRI, the link function.
Table~\ref{tab:commands} lists the estimators and the corresponding
\proglang{R} functions and \proglang{Stata} commands.

\begin{table}[t!]
\centering
\small
\begin{tabular}{>{\raggedright\arraybackslash}p{3.6cm}>{\raggedright\arraybackslash}p{5.2cm}>{\raggedright\arraybackslash}p{2.5cm}>{\raggedright\arraybackslash}p{2.6cm}}
\toprule
Estimator & Description & \pkg{OneSampleMR} & \pkg{ivonesamplemr} \\
\midrule
Additive SMM (ASMM) & Linear IV estimator; causal risk difference for a binary outcome & \fct{ivreg} from \pkg{ivreg} (see \code{?asmm}) & \code{ivregress} or \code{ivreg2} (see \code{help ivasmm}) \\
Multiplicative SMM (MSMM) & Causal risk ratio, with moment conditions \eqref{eq:msmm} or \eqref{eq:msmmalt} or two-stage least squares methods & \fct{msmm} & \code{ivmsmm} \\
Logistic SMM (LSMM) & Causal odds ratio among the exposed & \fct{lsmm} & \code{ivlsmm} \\
Two-stage predictor substitution (TSPS) & Second stage on the predicted exposure, with \code{identity}, \code{logadd}, \code{logmult}, or \code{logit} link & \fct{tsps} & \code{ivtsps} \\
Two-stage residual inclusion (TSRI) & Second stage on the exposure and first stage residuals, with the same links as TSPS & \fct{tsri} & \code{ivtsri} \\
Sanderson-Windmeijer conditional $F$-statistic & Instrument strength with multiple exposures & \fct{fsw} & \code{ivfsw} \\
Within-quantile estimation & Any of the above within quantiles of the first stage residuals & \fct{ivxtile} & \code{ivxtile:} \\
Moving window estimation & Any of the above within moving windows of the first stage residuals & \fct{ivmw} & \code{ivmw:} \\
\bottomrule
\end{tabular}
\caption{\label{tab:commands} Estimators and the corresponding functions in
\pkg{OneSampleMR} and commands in \pkg{ivonesamplemr}.}
\end{table}

%% -- Data ---------------------------------------------------------------------

\section{Simulated example data} \label{sec:data}

Individual-level data from MR studies, such as UK Biobank or birth cohort
studies, generally cannot be shared publicly. We therefore illustrate the
packages using a simulated dataset of $n = 10{,}000$ participants, generated by
the replication script \code{code.R} and saved as \code{mrdata.csv} so that
exactly the same data are analyzed in \proglang{R} and \proglang{Stata}. There
are three instruments, \code{z1}, \code{z2}, and \code{z3}, which mimic
single nucleotide polymorphisms coded 0, 1, 2; a binary measured confounder,
\code{w}, which affects both the exposures and the outcomes; an unmeasured confounder, $U \sim \text{Uniform}(0, 1)$; two
continuous exposures, \code{x} and \code{x2}; a binary outcome, \code{y}; and
a continuous outcome, \code{yc}. In the data generating model these variables
are denoted by the corresponding capital letters, for example $Z_1$ for
\code{z1} and $Y_c$ for \code{yc}:
\begin{align*}
  Z_1 &\sim \text{Bin}(2, 0.30), \quad Z_2 \sim \text{Bin}(2, 0.25), \quad
  Z_3 \sim \text{Bin}(2, 0.40), \\
  W &\sim \text{Bernoulli}(0.5), \quad U \sim \text{Uniform}(0, 1), \\
  X &= 0.5 Z_1 + 0.4 Z_2 + 0.3 Z_3 + 0.5 W + U + \varepsilon_1, \\
  X_2 &= 0.4 Z_1 + 0.1 Z_2 + 0.3 Z_3 + 0.3 W + U + \varepsilon_2, \\
  \Pr(Y = 1 \mid X, W, U) &= \exp\{-3 + \log(1.3) X + U + 0.3 W\}, \\
  Y_c &= 0.5 X + 0.3 X_2 + W + U + \varepsilon_3,
\end{align*}
where $\varepsilon_1, \varepsilon_2, \varepsilon_3$ are independent standard normal
errors. The binary outcome follows a log-linear model with no effect
modification of the exposure effect by the instruments, so that the MSMM is
correctly specified with a true causal risk ratio of $1.3$ per unit increase
in \code{x}. The continuous outcome has causal effects of $0.5$ and $0.3$ of
\code{x} and \code{x2}, respectively. About 16\% of participants have
$Y = 1$. In \proglang{R}, the first three rows of the simulated data are:

\begin{CodeChunk}
\begin{CodeInput}
R> library("OneSampleMR")
R> library("ivreg")
R> mrdata <- read.csv("mrdata.csv")
R> head(mrdata, 3)
\end{CodeInput}
\begin{CodeOutput}
  y    yc      x     x2 w z1 z2 z3
1 0 0.519 0.1012 1.1726 1  0  1  0
2 0 2.475 1.2201 0.4329 1  1  1  0
3 0 1.866 1.7331 0.8843 0  2  0  0
\end{CodeOutput}
\end{CodeChunk}

%% -- R package ----------------------------------------------------------------

\section[The OneSampleMR package for R]{The \pkg{OneSampleMR} package for \proglang{R}} \label{sec:r}

\pkg{OneSampleMR} can be installed from CRAN and loaded with
\begin{CodeChunk}
\begin{CodeInput}
R> install.packages("OneSampleMR")
R> library("OneSampleMR")
\end{CodeInput}
\end{CodeChunk}
The package exports seven functions: \fct{msmm}, \fct{lsmm}, \fct{tsps}, and
\fct{tsri}, which fit the estimators; \fct{fsw}, which computes the
conditional $F$-statistics; and \fct{ivxtile} and \fct{ivmw}, which fit an
estimator within quantile groups or moving windows of the first stage
residuals. The fitted models have \fct{print}, \fct{summary}, \fct{coef},
\fct{vcov}, \fct{confint}, \fct{nobs}, and \fct{formula} methods. It imports \pkg{gmm} \citep{Chausse:2010} for GMM
estimation, \pkg{ivreg} \citep{ivreg} for 2SLS, \pkg{Formula}
\citep{Zeileis+Croissant:2010} for multi-part formulas, \pkg{lmtest}
\citep{Zeileis+Hothorn:2002} for the Wald tests in \fct{fsw}, and \pkg{msm}
\citep{Jackson:2011} for delta method standard errors.

The estimation functions share the interface of \fct{ivreg} from \pkg{ivreg}.
Models are specified with a three-part formula, \code{y ~ x + w | z1 + z2 +
z3 + w}, in which the first part is the outcome, the second part lists the
exposure and any covariates, and the third part lists the instruments and the
same covariates. Variables appearing in both the second and third parts are
treated as exogenous covariates, and causal effect estimates are only reported
for the exposures. The functions also accept the \code{data},
\code{subset}, \code{na.action}, and \code{contrasts} arguments of
\fct{ivreg}, and starting values for the GMM optimizer can be supplied with
\code{t0}.

\subsection{Additive structural mean model}

As described in Section~\ref{sec:models}, the ASMM is fitted by any linear IV
estimator, and so \pkg{OneSampleMR} does not provide a separate function; its
\code{?asmm} help page directs users to \fct{ivreg}. For the binary outcome,
the estimated causal risk difference per unit increase in \code{x} is $0.035$
(95\% confidence interval, CI, $0.019$ to $0.051$).
\begin{CodeChunk}
\begin{CodeInput}
R> fit_asmm <- ivreg(y ~ x + w | z1 + z2 + z3 + w, data = mrdata)
R> summary(fit_asmm, diagnostics = TRUE)$diagnostics
\end{CodeInput}
\begin{CodeOutput}
                 df1  df2 statistic p-value
Weak instruments   3 9995  630.1997 0.00000
Wu-Hausman         1 9996    5.5271 0.01874
Sargan             2   NA    0.7591 0.68417
\end{CodeOutput}
\end{CodeChunk}
The first stage $F$-statistic of 630 indicates strong instruments.

\subsection{Multiplicative structural mean model}

\fct{msmm} fits the MSMM. Its \code{estmethod} argument selects between GMM
estimation of moment conditions \eqref{eq:msmm} (\code{"gmm"}, the default),
GMM estimation of the alternative moment conditions \eqref{eq:msmmalt}
(\code{"gmmalt"}), and, for binary $X$ and $Y$, the two 2SLS approaches of
\citet{Clarke+Palmer+Windmeijer:2015} (\code{"tsls"} and \code{"tslsalt"}).
The outcome must be a nonnegative integer, so that \fct{msmm} can be used for
binary and count outcomes. Multiple exposures are permitted with the GMM
methods.
\begin{CodeChunk}
\begin{CodeInput}
R> fit_msmm <- msmm(y ~ x | z1 + z2 + z3, data = mrdata)
R> summary(fit_msmm)
\end{CodeInput}
\begin{CodeOutput}
Call:
msmm(formula = y ~ x | z1 + z2 + z3, data = mrdata)

Estimation method: gmm

Coefficients:
        Estimate Std. Error z value Pr(>|z|)
E[Y(0)]  0.10947    0.00995   11.01  < 2e-16 ***
x        0.21181    0.05120    4.14  3.5e-05 ***
---
Signif. codes:  0 '***' 0.001 '**' 0.01 '*' 0.05 '.' 0.1 ' ' 1

Hansen J-test of the overidentifying restrictions:
J = 1.5 on 2 degrees of freedom, p-value: 0.472

E[Y(0)] with 95% CI:
      0.025 0.975
0.109 0.090 0.129

Causal risk ratio with 95% CI:
   CRR 0.025 0.975
x 1.24  1.12  1.37
\end{CodeOutput}
\end{CodeChunk}
The \fct{summary} method reports the parameter estimates with their standard
errors, Wald $z$ statistics, and $p$-values. The first parameter is the
estimated mean exposure-free potential outcome, $\E\{Y(0)\}$, and the second
the log causal risk ratio. The estimated causal risk ratio is $1.24$ (95\% CI
$1.12$ to $1.37$), consistent with the true value of $1.3$. The Hansen
$J$-test of the overidentifying restrictions is $1.50$ on 2 degrees of freedom
($p = 0.47$), giving no evidence against the validity of the instruments. The
\fct{print} method gives a shorter display of the estimates.

Covariates are adjusted for by including them in both parts of the formula.
This corresponds to a log-linear model for the exposure-free potential outcome
mean in the covariates, and is the same model as fitted by \code{ivpoisson} in
\proglang{Stata} with exogenous covariates. Here we use the alternative moment
conditions and adjust for \code{w}.
\begin{CodeChunk}
\begin{CodeInput}
R> fit_msmm_alt <- msmm(y ~ x + w | z1 + z2 + z3 + w, data = mrdata,
+    estmethod = "gmmalt")
R> fit_msmm_alt
\end{CodeInput}
\begin{CodeOutput}
Estimation method: gmmalt

Method
 twoStep

Objective function value:  0.0001872

log E[Y(0)]            x            w
   -2.35342      0.20110      0.29947

Convergence code =  0

E[Y(0)] when the covariates are 0 with 95% CI:
        0.025  0.975
0.0950 0.0802 0.1126

Causal risk ratio with 95% CI:
   CRR 0.025 0.975
x 1.22   1.1  1.36

Adjusted for: w
\end{CodeOutput}
\end{CodeChunk}
The causal risk ratio is only reported for the exposure, and $\E\{Y(0)\}$ is
now estimated at covariate values of 0.

\subsection{Logistic structural mean model}

\fct{lsmm} fits the double logistic SMM by two-step GMM. The association model
is a logistic regression of the outcome on the exposures, instruments, and
covariates, and the association and causal model moment conditions are fitted
jointly, using the same initial weight matrix, optimal weight matrix, and
starting values as \code{ivlsmm} in \pkg{ivonesamplemr}. Covariates are
included in both the association model and the causal model.
\begin{CodeChunk}
\begin{CodeInput}
R> fit_lsmm <- lsmm(y ~ x + w | z1 + z2 + z3 + w, data = mrdata)
R> fit_lsmm
\end{CodeInput}
\begin{CodeOutput}
Method
 twoStep

Objective function value:  7.2e-05

assoc:(Intercept)            assoc:x           assoc:z1
        -2.402243           0.426996          -0.080224
         assoc:z2           assoc:z3            assoc:w
        -0.062534          -0.051071           0.290310
                x                  w            E[Y(0)]
         0.265207           0.362480           0.090613

Convergence code =  0

E[Y(0)] when the covariates are 0 with 95% CI:
        0.025  0.975
0.0906 0.0748 0.1065

Causal odds ratio with 95% CI:
  COR 0.025 0.975
x 1.3  1.16  1.47

Adjusted for: w
\end{CodeOutput}
\end{CodeChunk}
The parameters prefixed \code{assoc:} belong to the association model, and
the coefficient of \code{x} in the causal model gives a causal odds ratio among
the exposed of $1.30$ (95\% CI $1.16$ to $1.47$). The \fct{summary} method
reports the Hansen $J$-test, here $0.72$ on 2 degrees of freedom ($p = 0.70$).

\subsection{Two-stage predictor substitution}

\fct{tsps} fits TSPS estimators, with the second stage link function chosen by
the \code{link} argument, one of \code{"identity"} (the default),
\code{"logadd"}, \code{"logmult"}, and \code{"logit"}. Here we adjust for the
covariate \code{w} in both stages and estimate a causal risk ratio with the
log-additive link.
\begin{CodeChunk}
\begin{CodeInput}
R> fit_tsps <- tsps(y ~ x + w | z1 + z2 + z3 + w, data = mrdata,
+    link = "logadd")
R> fit_tsps
\end{CodeInput}
\begin{CodeOutput}
Second stage model link function: logadd

Method
 twoStep

Objective function value:  1.905e-27

stage1:(Intercept)           stage1:z1           stage1:z2
           0.52301             0.49345             0.38829
         stage1:z3            stage1:w         (Intercept)
           0.31523             0.44639            -2.32105
                 x                   w
           0.21540             0.29672

Convergence code =  0

Estimates with 95% CI limits:
                   Estimate  0.025  0.975
stage1:(Intercept)    0.523  0.478  0.568
stage1:z1             0.493  0.462  0.525
stage1:z2             0.388  0.355  0.422
stage1:z3             0.315  0.286  0.345
stage1:w              0.446  0.406  0.487
(Intercept)          -2.321 -2.467 -2.175
x                     0.215  0.119  0.312
w                     0.297  0.197  0.397

Causal risk ratio with 95% CI limits:
  Estimate 0.025 0.975
x     1.24  1.13  1.37
\end{CodeOutput}
\end{CodeChunk}
The output reports all the parameters of the stacked GMM model. The parameters
prefixed \code{stage1:} are the first stage coefficients and the remaining
parameters are the second stage coefficients. Only the coefficient of the
exposure is exponentiated, giving an estimated causal risk ratio of $1.24$
(95\% CI $1.13$ to $1.37$).

\subsection{Two-stage residual inclusion}

\fct{tsri} has the same arguments as \fct{tsps}. Here we estimate a causal odds
ratio using the logit link.
\begin{CodeChunk}
\begin{CodeInput}
R> fit_tsri <- tsri(y ~ x + w | z1 + z2 + z3 + w, data = mrdata,
+    link = "logit")
R> fit_tsri
\end{CodeInput}
\begin{CodeOutput}
Second stage model link function: logit

Method
 twoStep

Objective function value:  1.04e-18

stage1:(Intercept)           stage1:z1           stage1:z2
           0.52301             0.49345             0.38829
         stage1:z3            stage1:w         (Intercept)
           0.31523             0.44639            -2.31417
                 x    stage1:residuals                   w
           0.26327             0.16315             0.36263

Convergence code =  0

Estimates with 95% CI limits:
                   Estimate   0.025  0.975
stage1:(Intercept)    0.523  0.4779  0.568
stage1:z1             0.493  0.4619  0.525
stage1:z2             0.388  0.3550  0.422
stage1:z3             0.315  0.2856  0.345
stage1:w              0.446  0.4056  0.487
(Intercept)          -2.314 -2.4929 -2.135
x                     0.263  0.1437  0.383
stage1:residuals      0.163  0.0322  0.294
w                     0.363  0.2410  0.484

Causal odds ratio with 95% CI limits:
  Estimate 0.025 0.975
x      1.3  1.15  1.47
\end{CodeOutput}
\end{CodeChunk}
The coefficient of \code{x} gives a causal odds ratio of $1.30$ (95\% CI
$1.15$ to $1.47$). As discussed in Section~\ref{sec:models}, this is a
conditional odds ratio, which differs from the causal risk ratio estimated by
the MSMM both because odds ratios exceed risk ratios for a common outcome and
because of noncollapsibility. The coefficient of the first stage residuals,
\code{stage1:residuals}, differs from zero (95\% CI $0.032$ to $0.294$),
which is evidence of confounding of the exposure-outcome association.

The \fct{coef}, \fct{vcov}, and \fct{confint} methods return the parameter
estimates, their variance-covariance matrix, and Wald confidence intervals on
the scale on which they are estimated, here the log odds ratio scale.
\begin{CodeChunk}
\begin{CodeInput}
R> confint(fit_tsri, "x")
\end{CodeInput}
\begin{CodeOutput}
   2.5 % 97.5 %
x 0.1437 0.3829
\end{CodeOutput}
\end{CodeChunk}
Exponentiating these limits gives the confidence interval for the causal odds
ratio shown above.

\subsection{Nonlinear exposure-outcome relationships}

\begin{figure}[t!]
\centering
\includegraphics[width=0.7\textwidth]{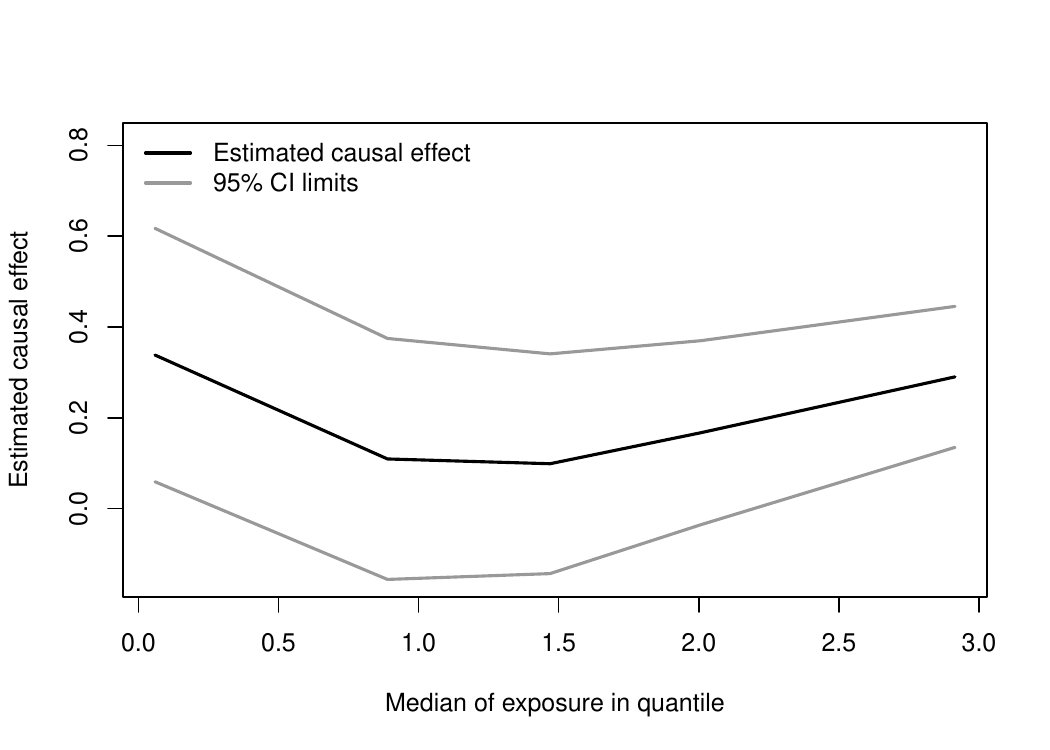}
\caption{\label{fig:ivxtile-r} Log causal risk ratio estimates (with 95\%
confidence intervals) from the MSMM fitted within quintiles of the first stage
residuals, plotted against the median exposure in each quintile, produced by
\code{plot(res\_xtile)}.}
\end{figure}

\fct{ivxtile} and \fct{ivmw} fit any of the estimators within quantile groups
of the first stage residuals or within moving windows of a given number of
participants ordered by the first stage residuals, respectively
(Section~\ref{sec:models}). The estimator is chosen by the \code{estimator}
argument, for example \fct{ivreg} (the default), \fct{msmm}, or \fct{tsps},
and further arguments are passed to it. The quantile groups are defined in the
same way as by \proglang{Stata}'s \code{xtile} command, so that the results
match those of \pkg{ivonesamplemr}. Here we fit the MSMM, with the alternative
moment conditions, within quintiles.
\begin{CodeChunk}
\begin{CodeInput}
R> res_xtile <- ivxtile(y ~ x | z1 + z2 + z3, data = mrdata, nq = 5,
+    estimator = msmm, estmethod = "gmmalt")
R> res_xtile
\end{CodeInput}
\begin{CodeOutput}
Estimates of x within quantiles of the first stage residuals:

 quantile    n  beta     se medexposure   lowci uppci
        1 2000 0.338 0.1424      0.0592  0.0593 0.617
        2 2000 0.109 0.1354      0.8888 -0.1559 0.375
        3 2000 0.099 0.1234      1.4692 -0.1430 0.341
        4 2000 0.167 0.1034      2.0052 -0.0358 0.370
        5 2000 0.290 0.0792      2.9143  0.1350 0.446
\end{CodeOutput}
\end{CodeChunk}
The output gives the log causal risk ratio, its standard error and 95\% CI, and
the median exposure in each quintile. The overlapping confidence intervals are
consistent with the constant causal risk ratio of the data generating model.
The \fct{plot} method plots the estimates against the median exposure
(Figure~\ref{fig:ivxtile-r}). Similarly, for example,
\code{ivmw(y ~ x | z1 + z2 + z3, data = mrdata, estimator = msmm,}
\code{window = 2000, step = 250)} fits the MSMM in windows of 2000 participants
starting every 250 participants.

\subsection{Conditional F-statistics}

\fct{fsw} computes the Sanderson-Windmeijer conditional $F$-statistics for an
IV model fitted by \fct{ivreg} from \pkg{ivreg} or \pkg{AER},
\fct{iv\_robust} from \pkg{estimatr}, or \fct{feols} from \pkg{fixest}. Here we
fit a multivariable MR model of the continuous outcome on both exposures.
\begin{CodeChunk}
\begin{CodeInput}
R> fit_two <- ivreg(yc ~ x + x2 + w | z1 + z2 + z3 + w, data = mrdata)
R> fsw(fit_two)
\end{CodeInput}
\begin{CodeOutput}
Model sample size:  10000

Sanderson-Windmeijer conditional F-statistics for first stage model:
   F value d.f. Residual d.f. Pr(>F)
x    63.24    2          9995 <2e-16 ***
x2   59.82    2          9995 <2e-16 ***
---
Signif. codes:  0 '***' 0.001 '**' 0.01 '*' 0.05 '.' 0.1 ' ' 1
\end{CodeOutput}
\end{CodeChunk}
Although the usual first stage $F$-statistics for \code{x} and \code{x2} are
$630$ and $340$, the conditional $F$-statistics are an order of magnitude
smaller, $63$ and $60$, because the instruments have similar associations with
both exposures. Following \citet{Sanderson+Windmeijer:2016}, the conditional
$F$-statistics can be compared with the critical values of
\citet{Stock+Yogo:2005} for a single exposure with $L - K + 1 = 2$
instruments. With 2 instruments only the critical values based on the size of
the Wald test are available. The 5\% critical value for a test of the null
hypothesis that the actual size of a nominal 5\% Wald test on the IV estimate
is more than 10\% is $19.93$. Both conditional $F$-statistics are well above
this, so we reject the null hypothesis that either exposure is weakly
identified. The 2SLS estimates, $0.447$ (standard error $0.058$) for \code{x} and $0.350$ ($0.078$)
for \code{x2}, are close to the true values of $0.5$ and $0.3$.

%% -- Stata package ------------------------------------------------------------

\section[The ivonesamplemr package for Stata]{The \pkg{ivonesamplemr} package for \proglang{Stata}} \label{sec:stata}

\pkg{ivonesamplemr} can be installed from the SSC archive, together with its
dependencies \code{ivreg2}, \code{ranktest}, and \code{ivpois}, with
\begin{CodeChunk}
\begin{CodeInput}
. foreach p in ivonesamplemr ranktest ivreg2 ivpois {
  2.   ssc install `p'
  3. }
\end{CodeInput}
\end{CodeChunk}
The package provides the commands listed in Table~\ref{tab:commands},
together with an overview help file (\code{help ivonesamplemr}), and an \code{iv} prefix so
that, for example, \code{ivlsmm} can also be called as \code{iv lsmm}. The
package requires \proglang{Stata}~13 or later, because \code{ivmsmm} uses
\code{ivpoisson}, which was introduced in \proglang{Stata}~13, and its
certification scripts have been run under \proglang{Stata}~17.0, 18.0,
StataNow~18.5, and StataNow~19.5.

All estimation commands use the syntax of \code{ivregress}:
\begin{Code}
command depvar [varlist1] (varlist2 = varlist_iv) [, options]
\end{Code}
where \code{varlist1} are the exogenous covariates, \code{varlist2} the
exposures, and \code{varlist\_iv} the instruments. We first read in the same
simulated data.
\begin{CodeChunk}
\begin{CodeInput}
. import delimited using mrdata.csv, clear
\end{CodeInput}
\begin{CodeOutput}
(8 vars, 10,000 obs)
\end{CodeOutput}
\end{CodeChunk}

\subsection{Structural mean models}

\code{ivmsmm} fits the MSMM and reports the causal risk ratio (the
\code{noirr} option suppresses exponentiation). Its \code{estmethod()}
option selects the same estimation methods as \fct{msmm}: by default,
\code{estmethod(gmmalt)}, it is a wrapper for \code{ivpoisson} and fits the
moment conditions in \eqref{eq:msmmalt}; \code{estmethod(gmm)} fits the
moment conditions in \eqref{eq:msmm} using \code{gmm}; and
\code{estmethod(tsls)} and \code{estmethod(tslsalt)} fit the two-stage least
squares methods for a binary exposure and outcome using \code{ivregress}.
Further options are passed to these commands. Here we use the default method
and adjust for \code{w}.
\begin{CodeChunk}
\begin{CodeInput}
. ivmsmm y w (x = z1 z2 z3)
\end{CodeInput}
\begin{CodeOutput}
Final GMM criterion Q(b) = .0001887

Exponential mean model with endogenous regressors

Number of parameters =   3                         Number of obs  =     10,000
Number of moments    =   5
Initial weight matrix: Unadjusted
GMM weight matrix:     Robust

------------------------------------------------------------------------------
             |               Robust
           y |        IRR   std. err.      z    P>|z|     [95% conf. interval]
-------------+----------------------------------------------------------------
           x |   1.222707   .0649926     3.78   0.000     1.101735    1.356963
           w |   1.349156   .0725141     5.57   0.000     1.214261    1.499037
       _cons |   .0950487   .0082171   -27.22   0.000     .0802341    .1125987
------------------------------------------------------------------------------
Note: _cons estimates baseline incidence rate.
Endogenous: x
Exogenous:  w z1 z2 z3
\end{CodeOutput}
\end{CodeChunk}
The causal risk ratio for \code{x} matches that from
\code{msmm(..., estmethod = "gmmalt")} in Section~\ref{sec:r}. The row for
\code{w} is the exponentiated covariate coefficient of the model for the
exposure-free potential outcome, not a causal effect.
\proglang{Stata}'s \code{estat overid} postestimation command reports the
Hansen $J$-test.

\code{ivlsmm} fits the double logistic SMM. By default the association model
contains the exposure, instruments, and covariates; this can be changed with
the \code{amxb()} option.
\begin{CodeChunk}
\begin{CodeInput}
. ivlsmm y w (x = z1 z2 z3)
\end{CodeInput}
\begin{CodeOutput}
  (output omitted)
------------------------------------------------------------------------------
             |               Robust
             | Coefficient  std. err.      z    P>|z|     [95% conf. interval]
-------------+----------------------------------------------------------------
       /xb_x |   .4269958    .027912    15.30   0.000     .3722892    .4817024
      /xb_z1 |  -.0802244   .0332937    -2.41   0.016    -.1454789     -.01497
      /xb_z2 |  -.0625341   .0257154    -2.43   0.015    -.1129353   -.0121329
      /xb_z3 |  -.0510708   .0210212    -2.43   0.015    -.0922716   -.0098701
       /xb_w |   .2903099   .0570288     5.09   0.000     .1785354    .4020843
         /b0 |  -2.402243   .0687134   -34.96   0.000    -2.536919   -2.267567
     /cmxb_x |   .2652068   .0607984     4.36   0.000     .1460441    .3843694
     /cmxb_w |   .3624805    .061985     5.85   0.000     .2409921    .4839688
        /ey0 |   .0906126   .0080846    11.21   0.000      .074767    .1064582
------------------------------------------------------------------------------
Instruments for equation 1: x z1 z2 z3 w _cons
Instruments for equation 2: z1 z2 z3 w _cons

  Test of overidentifying restriction:

  Hansen's J chi2(2) = .719958 (p = 0.6977)

Causal odds ratio for: x

 ( 1)  [cmxb_x]_cons = 0

------------------------------------------------------------------------------
             |     exp(b)   Std. err.      z    P>|z|     [95% conf. interval]
-------------+----------------------------------------------------------------
         (1) |     1.3037   .0792629     4.36   0.000     1.157247    1.468688
------------------------------------------------------------------------------
\end{CodeOutput}
\end{CodeChunk}
The \code{xb\_} and \code{b0} parameters belong to the association model, and
\code{cmxb\_x} is the log causal odds ratio of the causal model, giving a
causal odds ratio among the exposed of $1.30$ (95\% CI $1.16$ to $1.47$). The
estimates match those from \fct{lsmm} in Section~\ref{sec:r}, where the
association model parameters are labelled \code{assoc:} and \code{cmxb\_w}
is the coefficient of \code{w} in the causal model.

\subsection{Two-stage predictor substitution and residual inclusion}

\code{ivtsps} and \code{ivtsri} take the \code{link()} option with the same
four link functions as \pkg{OneSampleMR}. The causal effect is the parameter
\code{b1}, which is reported for the exposure, exponentiated for the nonlinear
links, after the coefficient table. The \code{estonly}
option fits the two regressions without the subsequent GMM step, which is
useful with the \code{bootstrap} prefix.
\begin{CodeChunk}
\begin{CodeInput}
. ivtsri y w (x = z1 z2 z3), link(logit)
\end{CodeInput}
\begin{CodeOutput}
Final GMM criterion Q(b) = 1.62e-33

note: model is exactly identified.

GMM estimation

Number of parameters =   9
Number of moments    =   9
Initial weight matrix: Unadjusted                 Number of obs   =     10,000

------------------------------------------------------------------------------
             |               Robust
             | Coefficient  std. err.      z    P>|z|     [95% conf. interval]
-------------+----------------------------------------------------------------
    /s1xb_z1 |   .4934457   .0161039    30.64   0.000     .4618827    .5250088
    /s1xb_z2 |   .3882909   .0170055    22.83   0.000     .3549609     .421621
    /s1xb_z3 |   .3152321   .0151272    20.84   0.000     .2855834    .3448809
     /s1xb_w |   .4463881    .020786    21.48   0.000     .4056482     .487128
         /a0 |   .5230142   .0230039    22.74   0.000     .4779273     .568101
         /b1 |   .2632662   .0610221     4.31   0.000      .143665    .3828673
         /b2 |   .1631531   .0668092     2.44   0.015     .0322095    .2940966
 /s2exogxb_w |   .3626276   .0620558     5.84   0.000     .2410005    .4842547
         /b0 |  -2.314169   .0912082   -25.37   0.000    -2.492934   -2.135404
------------------------------------------------------------------------------
Instruments for equation 1: z1 z2 z3 w _cons
Instruments for equation 2: x residuals w _cons

Causal odds ratio for: x

 ( 1)  [b1]_cons = 0

------------------------------------------------------------------------------
             |     exp(b)   Std. err.      z    P>|z|     [95% conf. interval]
-------------+----------------------------------------------------------------
         (1) |   1.301173   .0794003     4.31   0.000     1.154497    1.466483
------------------------------------------------------------------------------
\end{CodeOutput}
\end{CodeChunk}
The \code{s1xb\_} and \code{a0} parameters are the first stage coefficients,
\code{b1} is the causal effect, \code{b2} the coefficient of the first stage
residuals (which is not a causal effect and so is not exponentiated), and \code{s2exogxb\_w} and \code{b0} the remaining second stage
coefficients. The estimates are identical to those from \fct{tsri} in
Section~\ref{sec:r}.

\subsection{Conditional F-statistics}

\code{ivfsw} computes the Sanderson-Windmeijer conditional $F$-statistics
after a model fitted by \code{ivregress} or \code{ivreg2}, taking the
exposures, covariates, and instruments from the estimation results.
\begin{CodeChunk}
\begin{CodeInput}
. quietly ivregress 2sls yc w (x x2 = z1 z2 z3)
. ivfsw
\end{CodeInput}
\begin{CodeOutput}
Sanderson-Windmeijer conditional F-statistics
Number of obs = 10000

Variable                F      df      df_r       P>F
------------------------------------------------------
x                   63.24       2      9995    0.0000
x2                  59.82       2      9995    0.0000
\end{CodeOutput}
\end{CodeChunk}
These match the output of \fct{fsw} in Section~\ref{sec:r}, and the
Sanderson-Windmeijer $F$-statistics reported by \code{ivreg2} with the
\code{first} option.

\subsection{Nonlinear exposure-outcome relationships}

The \code{ivxtile} and \code{ivmw} prefix commands apply any of
\code{ivregress}, \code{ivreg2}, \code{ivmsmm}, \code{ivlsmm}, \code{ivtsps},
or \code{ivtsri} within quantiles of the first stage residuals, or within
moving windows of a given number of participants ordered by the first stage
residuals, respectively. The \code{par()} option names the parameter to
collect, and the \code{saving()} option saves the results to a dataset.
Both commands also plot the estimates against the median exposure in each
quantile or window. Here we fit the MSMM within quintiles; the plot is
equivalent to Figure~\ref{fig:ivxtile-r}.
\begin{CodeChunk}
\begin{CodeInput}
. ivxtile, nq(5) par(x): ivmsmm y (x = z1 z2 z3)
\end{CodeInput}
\begin{CodeOutput}
    quantile       beta         se   medendog       lowci      uppci
           1     .33837   .1423763   .0592363    .0593123   .6174276
           2   .1091023   .1353864   .8888213    -.156255   .3744597
           3   .0988285   .1234447   1.469166   -.1431231   .3407801
           4   .1669726   .1034489   2.005172   -.0357872   .3697325
           5   .2901259   .0792011   2.914309    .1348917   .4453601
\end{CodeOutput}
\end{CodeChunk}
The output gives the log causal risk ratio, its standard error and 95\% CI,
and the median exposure (\code{medendog}) in each quintile. The estimates
match those of \fct{ivxtile} in Section~\ref{sec:r} to three decimal places;
the small differences are due to the different weight matrices used by
\code{ivpoisson} and the \pkg{gmm} package in the second step of the GMM
estimation. For the moving window approach, for example
\code{ivmw, window(2000) stepsize(250) par(x): ivmsmm y (x = z1 z2 z3)} fits
the MSMM in windows of 2000 participants starting every 250 participants.

%% -- Agreement ----------------------------------------------------------------

\section{Agreement across implementations} \label{sec:agreement}

Table~\ref{tab:agreement} compares the estimates of the causal effect of
\code{x} on \code{y} from the two packages, all adjusted for \code{w}. The
estimates were produced by the comparison sections of the replication scripts
\code{code.R} and \code{code.do}. For every estimator, the estimates and
confidence limits agree to three decimal places, as do the
Sanderson-Windmeijer $F$-statistics and the within-quantile estimates in
Sections~\ref{sec:r} and \ref{sec:stata}.

The risk ratio estimates are all somewhat below the true value of $1.3$,
although their confidence intervals include it. This is due to sampling
variability in this dataset of 10,000 participants rather than to bias. In a
dataset of 100,000 participants simulated from the same model (see
\code{code.R}), the MSMM estimates were $1.318$ and $1.315$, the TSPS and TSRI
log-additive estimates $1.303$ and $1.304$, and the ASMM risk difference
estimate $0.044$.

\begin{table}[t!]
\centering
\small
\begin{tabular}{lll}
\toprule
Estimator & \pkg{OneSampleMR} & \pkg{ivonesamplemr} \\
\midrule
ASMM (2SLS), risk difference          & 0.035 (0.019, 0.051) & 0.035 (0.019, 0.051) \\
MSMM, moment conditions \eqref{eq:msmm}, risk ratio    & 1.233 (1.110, 1.369) & 1.233 (1.110, 1.369) \\
MSMM, moment conditions \eqref{eq:msmmalt}, risk ratio & 1.223 (1.102, 1.357) & 1.223 (1.102, 1.357) \\
LSMM, odds ratio                      & 1.304 (1.157, 1.469) & 1.304 (1.157, 1.469) \\
TSPS, identity, risk difference       & 0.035 (0.019, 0.051) & 0.035 (0.019, 0.051) \\
TSPS, log-additive, risk ratio        & 1.240 (1.127, 1.366) & 1.240 (1.127, 1.366) \\
TSPS, log-multiplicative, risk ratio  & 1.226 (1.110, 1.353) & 1.226 (1.110, 1.353) \\
TSPS, logit, odds ratio               & 1.297 (1.154, 1.457) & 1.297 (1.154, 1.457) \\
TSRI, identity, risk difference       & 0.035 (0.019, 0.051) & 0.035 (0.019, 0.051) \\
TSRI, log-additive, risk ratio        & 1.235 (1.123, 1.359) & 1.235 (1.123, 1.359) \\
TSRI, log-multiplicative, risk ratio  & 1.231 (1.104, 1.373) & 1.231 (1.104, 1.373) \\
TSRI, logit, odds ratio               & 1.301 (1.154, 1.466) & 1.301 (1.154, 1.466) \\
\bottomrule
\end{tabular}
\caption{\label{tab:agreement} Estimates (95\% CI) of the causal effect of
\code{x} on \code{y}, adjusted for \code{w}, from \pkg{OneSampleMR} and
\pkg{ivonesamplemr} applied to the same simulated data. The data were
generated with a constant causal risk ratio of $1.3$, which implies an average
marginal causal risk difference per unit increase in \code{x} of $0.042$.
Because the odds ratio is noncollapsible and is not constant under this model,
there is no single true causal odds ratio.}
\end{table}

We also compared \fct{msmm} with G-estimation of the MSMM in \pkg{ivtools}.
Because \fct{ivglm}'s G-estimation uses a model for a single instrument, we
used the unweighted allele score \code{z1 + z2 + z3} as the instrument in both
packages:
\begin{CodeChunk}
\begin{CodeInput}
R> mrdata$score <- with(mrdata, z1 + z2 + z3)
R> fitZ <- glm(score ~ 1, data = mrdata)
R> fit_g <- ivtools::ivglm(estmethod = "g", X = "x", Y = "y",
+    fitZ.L = fitZ, data = mrdata, link = "log")
R> fit_m <- msmm(y ~ x | score, data = mrdata)
R> rbind(ivtools = exp(c(fit_g$est,
+    fit_g$est + c(-1, 1) * qnorm(0.975) * sqrt(diag(fit_g$vcov)))),
+    OneSampleMR = fit_m$crrci[1, ])
\end{CodeInput}
\begin{CodeOutput}
                x
ivtools     1.221 1.105 1.35
OneSampleMR 1.221 1.105 1.35
\end{CodeOutput}
\end{CodeChunk}
With a single instrument the model is exactly identified and both packages
solve the same estimating equation, $\E\{(Z - \E Z) Y \exp(-\psi X)\} = 0$.
The estimates of $\psi$ from \pkg{ivtools}, \fct{msmm}, and the root of the
estimating equation found by \fct{uniroot} agree to seven decimal places
($0.1999973$).

%% -- Summary/conclusions/discussion -------------------------------------------

\section{Summary and discussion} \label{sec:summary}

We have described \pkg{OneSampleMR} for \proglang{R} and \pkg{ivonesamplemr}
for \proglang{Stata}, which implement IV estimators for one-sample MR analyses
with the same functionality, parameterizations, and link function names.
The packages do not introduce new methodology. Rather, they make established
estimators, which are particularly relevant for binary outcomes, easy to apply
with correct standard errors and an interface familiar to users of
\fct{ivreg} in \proglang{R} and \code{ivregress} in \proglang{Stata}. Because
the two packages give the same results, analyses can be checked across
languages, and collaborators using different languages can fit the same
models.

Both packages provide the MSMM with both sets of moment conditions and the
two-stage least squares methods, the LSMM, the TSPS and TSRI estimators with
four link functions, the Sanderson-Windmeijer conditional $F$-statistic, and
within-quantile and moving window estimation, while the ASMM is fitted by the
standard linear IV commands, \fct{ivreg} and \code{ivregress}. A limitation of
both packages is that the TSPS and TSRI estimators allow only a single
exposure, and future work includes allowing multiple exposures in these
estimators.

Users should bear in mind that all the estimators rely on the IV assumptions,
which cannot be fully tested, and on additional modeling assumptions which
differ between estimators. The estimators also target different estimands: a
causal risk difference (ASMM, and TSPS and TSRI with the identity link), a
causal risk ratio (MSMM, and TSPS and TSRI with the log-additive and
log-multiplicative links), a causal odds ratio among the exposed (LSMM), and a
conditional causal odds ratio (TSPS and TSRI with the logit link). In our
example the risk ratio estimators agree closely because the data were generated
from a model in which their assumptions hold, with a constant causal risk ratio,
a linear first stage, and no effect modification by the instruments. When the
assumptions of an estimator do not hold, for example when the causal effect is
modified by the instruments or the first stage model is misspecified, estimators
that rely on different assumptions can give different results
\citep{Palmer+Sterne+Harbord:2011, Clarke+Palmer+Windmeijer:2015}. Fitting
several estimators, as facilitated by the packages, and comparing their
results, together with the overidentification tests when multiple instruments
are available, can therefore help assess the sensitivity of conclusions to these
assumptions.

%% -- Optional special unnumbered sections -------------------------------------

\section*{Computational details}

The results in this paper were obtained using \proglang{R}~4.6.2 with
\pkg{OneSampleMR}~0.2.0, \pkg{gmm}~1.9-1, \pkg{ivreg}~0.6-8, and
\pkg{ivtools}~2.3.0, and StataNow~19.5 MP with \pkg{ivonesamplemr}~0.2.0
(distribution date 23 November 2026), \code{ivreg2}~4.1.12,
\code{ranktest}~2.0.04, and \code{ivpois}~1.1.8. \proglang{R} itself and all \proglang{R} packages
used are available from CRAN at \url{https://CRAN.R-project.org/}, and the
\proglang{Stata} packages used are available from the SSC archive. The
replication materials consist of \code{code.R}, which simulates the data and
produces the \proglang{R} results, and \code{code.do}, which produces the
\proglang{Stata} results.

\section*{Acknowledgments}

% TODO: funding statements, e.g., MRC Integrative Epidemiology Unit funding.
TP acknowledges funding from NIHR510725. TP acknowledges support from MC\_UU\_00032/2. TP, ES, and NV acknowledge support from MC\_UU\_00032/1. ES and NV acknowledge
funding from the Medical Research Council UKRI077.

%% -- Bibliography -------------------------------------------------------------

\bibliography{refs}

\end{document}